\documentclass[runningheads]{llncs}
\newcommand{\ie}{\emph{i.e.}}
\newcommand{\eg}{\emph{e.g.}}
\usepackage[T1]{fontenc}
\usepackage{graphicx}
\usepackage{subfigure}

\begin{document}
\title{A Novel DID Method Leveraging the IOTA Tangle and its Integration into OpenSSL}
\titlerunning{A Novel DID Method Leveraging the IOTA Tangle}
\author{Alessio Claudio%
\and
Andrea Vesco %
}
\authorrunning{A.Claudio and A.Vesco}
\institute{LINKS Foundation - Cybersecurity Research Group \\
Torino 10138, Italy \\
\email{\{alessio.claudio, andrea.vesco\}@linksfoundation.com}}
\maketitle              %
\begin{abstract}
This paper presents, for the first time, a novel DID Method called Over-The-Tangle and discusses its design and working principles that leverage the IOTA Tangle as the Root-of-Trust for identity data. 
The results of a long lasting experimental test campaign in real-world settings suggests the adoption of a private gateway node synchronized with the IOTA Tangle on the mainnet for efficient DID control. Moreover, the paper
promotes the integration of the Decentralized IDentifier technology into OpenSSL through the use of Providers. A novel DID Operation and Provider is presented as a solution for building DID Method agility in OpenSSL.
\end{abstract}

\section{Introduction}
\label{introduction}
\begin{figure}[t]
    \begin{center}
    \includegraphics[width=8.5cm]{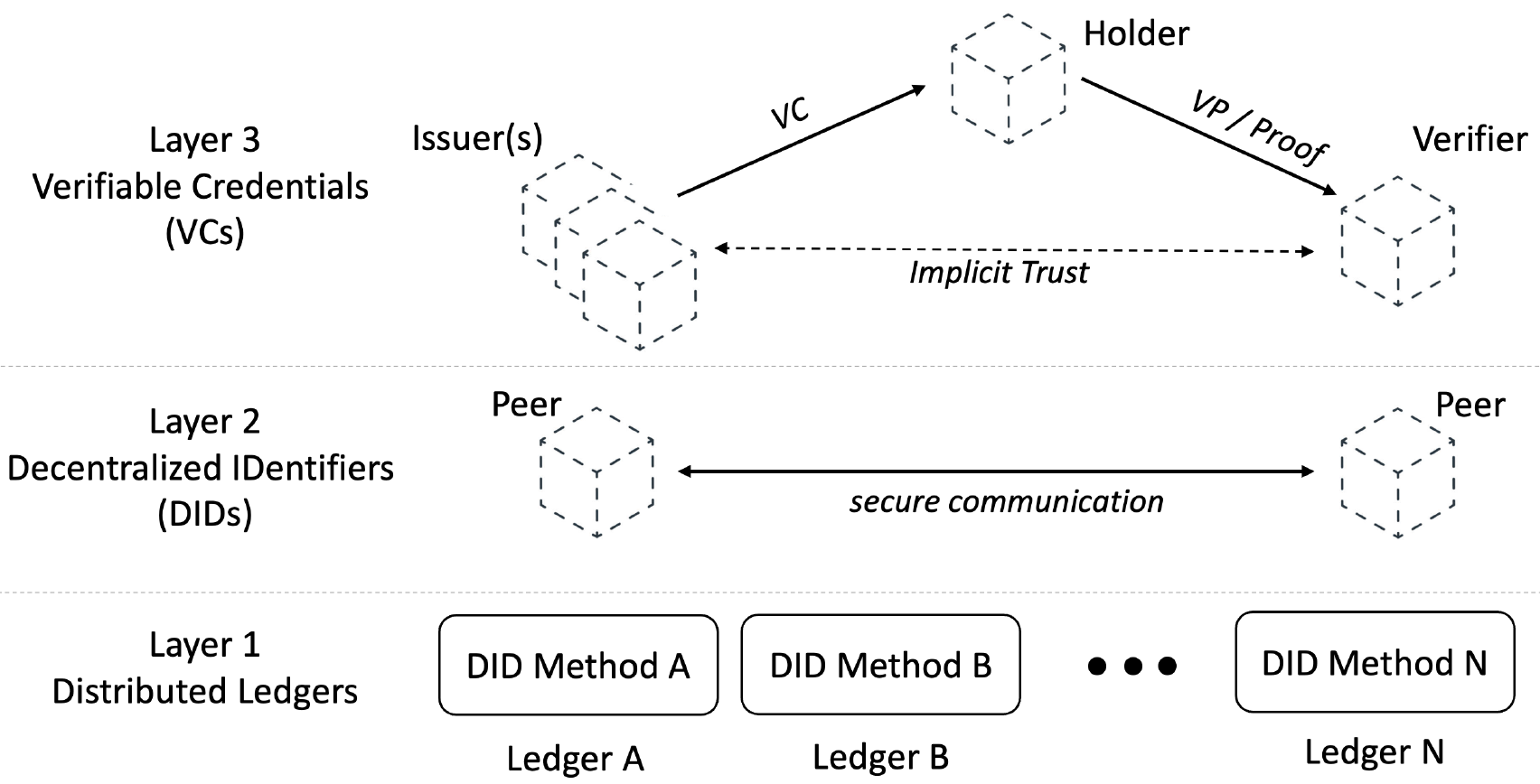}
    \end{center}
    \vspace{-0.5cm}
    \caption{The Self-Sovereign Identity stack.} 
    \label{ssi}
    \vspace{-0.5cm}
\end{figure}

The Self-Sovereign Identity (SSI)~\cite{SSI-book} is a decentralized digital identity paradigm that gives both human beings and things full control over the data they use to build and to prove their identity. The overall SSI stack, depicted in Fig.~\ref{ssi}, enables a new model for trusted digital interactions.

The Layer 1 is implemented by means of any Distributed Ledger Technology (DLT) acting as the Root-of-Trust (RoT) for identity data. In fact, DLTs are distributed immutable means of storage by design~\cite{DLTs}. A Decentralized IDentifier (DID)~\cite{DID} is a globally unique identity designed to verify a subject. The DIDs are Uniform Resource Identifiers (URIs) in the form \emph{did:method-name:method-specific-id} where \emph{method-name} is the name of the DID Method used to interact with the DLT and \emph{method-specific-id} is the pointer to the DID Document stored on the DLT. Thus, DIDs associate a DID subject with a DID Document~\cite{DID} to enable trustable interactions with that subject. The DID Method~\cite{DID,DID-registry} is a software implementation to interact with a specific ledger technology. In accordance with W3C recommendation~\cite{DID}, a DID Method must provide the primitives (\emph{i}) to create the DID, that is, generate a key pair ($sk_{id},pk_{id}$) for authentication purposes, the corresponding DID Document containing the public key of the pair and store the DID Document into the ledger at the \emph{method-specific-id} pointed by the DID, (\emph{ii}) to resolve a DID, that is, retrieve the DID Document from the \emph{method-specific-id} on the ledger pointed to by the DID and verify the validity of the DID, (\emph{iii}) to update the DID, that is, generate a new DID and corresponding DID Document while revoking the previous one and (\emph{iv}) to revoke a DID, that is, provide an immutable evidence on the ledger that a DID has been revoked by the owner of the DID. The DID Method implementation is ledger-specific and it makes the upper layers independent from the specific ledger technology.

The Layer 2 makes use of DIDs and DID Documents to establish a secure channel between two peers. In principle, both peers prove the ownership of the private key $sk_{id}$ bound to the public key $pk_{id}$ in the DID Document stored on the ledger. While the Layer 2 leverages DID technology (\ie\ the security foundation of the SSI stack) to begin the authentication procedure, the Layer 3 finalizes it and deals with authorization to services and to resources by means of Verifiable Credentials~\cite{VC}. VCs are secure and machine verifiable digital credentials. A VC is an unforgeable digital document that contains further characteristics of the digital identity of a peer than a simple key pair, a DID and the related DID Document. The addition of digital signatures makes VCs tamper-evident and trustworthy. 

\emph{The combination of the key pair ($sk_{id},pk_{id}$), a DID, the related DID Document and at least one VC forms the digital identity in the SSI world.} This composition of the digital identity reflects the decentralized nature of SSI. There is no authority that provides all the components of the identity to a peer and no authority is able to revoke completely the identity of a peer. Moreover, a peer can enrich its identity with multiple VCs issued by different issuers.
 
The Layer 3 works in accordance with the Triangle-of-Trust depicted in Fig.~\ref{ssi}. Three different roles coexist:
\begin{itemize}
  \item \textbf{Holder} is the peer that possesses one or more VCs and that generates a Verifiable Presentation (VP) to request a service or a resource from a Verifier; 
  \item \textbf{Issuer} is the peer that asserts claims about a subject, creates a VC from these claims, and issuess the VC to the Holders. Multiple Issuers can coexist.
  \item \textbf{Verifier} is the peer that receives a VP from the Holder and verifies the two signatures made by the Issuer on the VC and by the Holder on the VP before granting him/her access to a service or a resource based on the claims.
\end{itemize}

A VC contains the metadata to describe properties of the credential (\eg\ context, ID, type, Issuer of the VC, issuance and expiration dates) and most importantly, the claims about the identity of the subject in the \verb+credentialSubject+ field. %
The digital signature is made by the Issuer to make the VC an unforgeable and verifiable digital document. The Holder requests access to services and/or resources from the Verifier by presenting a VP. A VP is built as an envelop of the VC issued by an Issuer where signature is made by the Holder. Issuers are also responsible for VCs revocation for cryptographic integrity and for status change purposes~\cite{VC}. On top of these three layers, it is possible to build any ecosystem of trustable interactions among human beings and things.

This paper focuses on Layer 1 of the SSI stack and presents for the first time a novel DID Method called Over-The-Tangle (OTT). OTT works directly on top of the IOTA Tangle~\cite{Tangle}, a DLT offering feeless transactions to attach any type of data to the structure of the distributed ledger. The paper provides the following novel contributions: (\emph{i}) design and implementation of the OTT DID Method and (\emph{ii}) statistically relevant results of the overall performance of the solution in real world settings. Then, supported by the results achieved, the paper discusses some future perspectives and proposes to work on the design of a Transport Layer Security (TLS) handshake making use of DIDs and DID Documents while maintaining the interoperability with public key certificates.
As first step toward this objective, the paper provides two further contributions, (\emph{iii}) design and  implementation of the logic in OpenSSL~\cite{Openssl} for agile integration of DID Methods through a novel DID Provider. A Provider, in OpenSSL term, is a unit of code that provides one or more implementations for various operations for diverse algorithms that one might want to perform~\cite{Openssl-provider}; in the context of this work, the Provider collects a series of DID Methods and make them available to LibSSL, a sub-module of OpenSSL implementing the TLS handshake; this novel DID Operation universally defines the templates (\ie\ inputs and return values) for the DID Method functions \emph{Create}, \emph{Resolve}, \emph{Update}, and \emph{Revoke}, and (\emph{iv}) an easy to follow procedure to add other DID Methods in an agile fashion and open the way for further adoption of the solution.

\section{A DID Method for the IOTA Tangle}
OTT is a novel DID Method that leverages the IOTA Tangle as the RoT for DID Documents. The main working principles of the IOTA Tangle are detailed in~\cite{Tangle}. OTT is designed to work on top of the Chrysalis version of the IOTA protocol running on the IOTA mainnet. OTT makes use of indexation features of the IOTA Tangle. A peer can issue a transaction to attach a data to the Tangle while associating the data with a string 32 byte long. The Tangle indexes the strings and allows any peer to search for a data based on this string value; for this reason the string is called \emph{index}. Recalling the DID form and DID Document purpose detailed in Section~\ref{introduction}, in OTT the DID has the form \emph{did:ott:index} and a DID Document containing the public key $pk_{id}$ has the following structure:
\\
\begin{verbatim}
 "@context": ["https://www.w3.org/ns/did/v1"],
 "id":"did:ott:index",
 "created": "---Date-Hour---",
 "authenticationMethod": {
    "id": "did:ott:index#keys-0",
    "type": "---Verification-Method-Type---",
    "controller": "did:ott:index",
    "publicKeyPem": "---Public-Key---" }
\end{verbatim}

\subsection{Working principles}
\begin{figure}[h]
    \vspace{-0.5cm}
    \begin{center}
    \includegraphics[width=5.5cm]{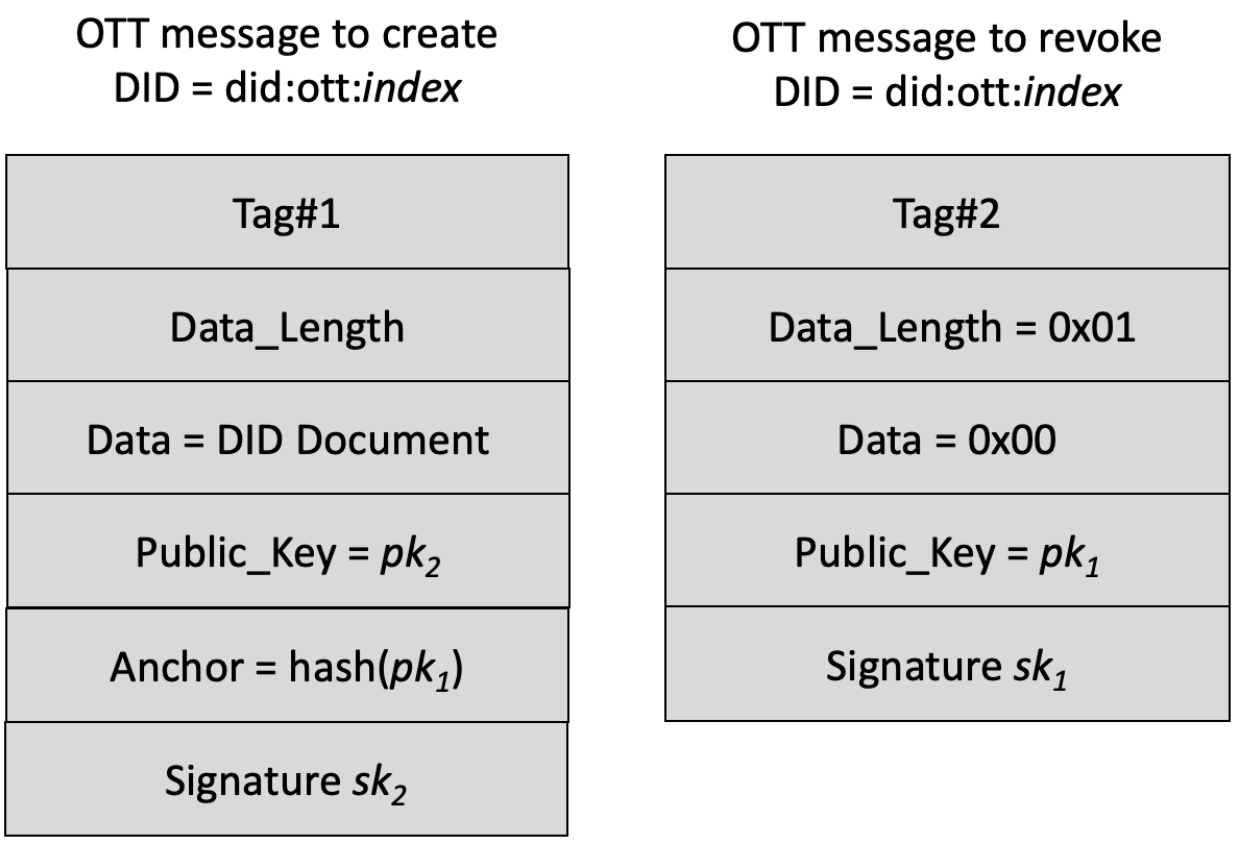}     
    \end{center}
    \vspace{-0.5cm}
    \caption{Structure of the OTT create message (left) and revoke message (right) that OTT uses in the \emph{Create, Resolve, Update} and \emph{Revoke} functions.} 
    \label{message}
    \vspace{-0.5cm}
\end{figure}

OTT uses two different messages to control a DID on the IOTA Tangle, see~Fig.~\ref{message}. The create message (left) and the revoke message (right) to create and revoke a DID respectively. An OTT message wraps the DID Document and it represents the unit of data attached to the Tangle through a single transaction. OTT also associates an \emph{index} to the messages for making them searchable. 

\begin{itemize}
  \item \textbf{Tag\#1} and \textbf{Tag\#2} [32 byte]: to identify OTT messages;
  \item \textbf{Data\_Length} [2 byte]: the length of the Data field in byte;
  \item \textbf{Data} [up to 31600 byte]: the DID Document or 0x00 in the case of an OTT revoke message;
  \item \textbf{Public\_Key} [32 byte]: ephemeral public key instrumental to establish the cryptographic binding between the create and the revoke messages associated with the same \emph{index};
  \item \textbf{Anchor} [32 byte] (only in create messages): the value instrumental to verify the cryptographic binding between the create and the revoke messages associated with the same \emph{index}; 
  \item \textbf{Signature} [64 byte]: digital signature of the message only for integrity purpose; made with Edwards-curve Digital Signature Algorithm (EdDSA) \cite{rfc8032} over the edwards25519 curve.
\end{itemize}

\begin{figure}[t]
    \begin{center}
    \includegraphics[width=9cm]{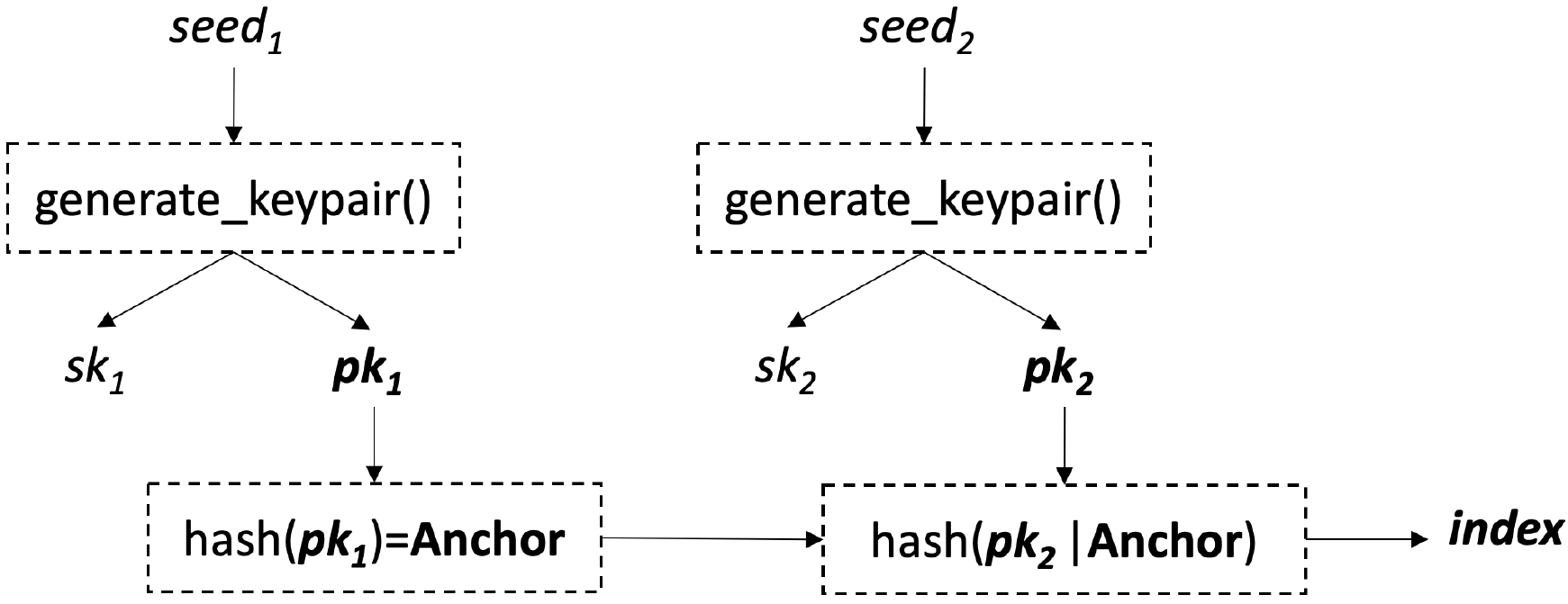} 
    \end{center}
    \vspace{-0.5cm}
    \caption{Generation of the \emph{index} and the ephemeral key pairs.} 
    \label{index}
    \vspace{-0.5cm}
\end{figure}

Fig.~\ref{index} depicts the process for generating two ephemeral key pairs ($sk_1,pk_1$) and ($sk_2,pk_2$) instrumental to generate the \emph{index} that OTT associates to create and revoke messages before attaching them to the Tangle. The process starts from two random seeds ($seed_1$ and $seed_2$) to generate the two ephemeral elliptic key pairs ($sk_1,pk_1$) and ($sk_2,pk_2$) over the edwards25519 curve. Then, the public key ($pk_2$) is concatenated with the digest, computed with the hash~\cite{rfc7693}, of the first public key ($pk_1$). The resulting value is finally hashed to generate the \emph{index}. The two ephemeral key pairs are bound to the value of \emph{index} and anyone who proves to own the first key pair also proves to own the second key pair.
Moreover, also knowing the \emph{index} and the Anchor value, the derivation of $pk_1$ is impracticable because the hash function is assumed to be not reversible. This construction is designed to ensure that (\emph{i}) the DID has been generated by the peer in accordance with OTT principles and (\emph{ii}) only that peer is able to revoke its own DID. In a permissionless DLT anyone can build and attach an OTT message to the Tangle and associate it with a previously selected DID, but only the owner, in OTT terminology, of the DID can build a valid OTT revoke message revealing the right $pk_1$ value.    
For the sake of clarity, the purposes of key pair ($sk_{id},pk_{id}$) and the key pairs ($sk_1, pk_1$), ($sk_2,pk_2$) are different. The first key pair is a piece of the digital identity of the peer, whereas the second couple of key pairs are ephemeral keys at the core of OTT working principles.

\vspace{-4mm}
\subsection{Implementation of the OTT Method functions}

OTT has been implemented leveraging the C language IOTA client library~\cite{IOTAC} and Libsodium cryptographic library~\cite{LIBSODIUM}. The IOTA client library provides the primitives to attach and retrieve messages to/from the Tangle. The primitive \emph{send\_indexation\_msg}() and the \emph{indexation payload} to attach a DID Document to the Tangle associating it with the specific DID URI, \ie\ \emph{did:ott:index}. The primitive \emph{find\_message\_by\_index}() to retrieve the DID Document from a given DID. OTT provides the implementation of the four functions to create, update and revoke DIDs and to resolve DIDs.

\begin{itemize}
  
  \item \textbf{Create} function is responsible for generating a new DID; this means generating an \emph{index} as in Fig.~\ref{index} and the related DID Document to be attached to the IOTA Tangle. This function encapsulate the DID Document into an OTT create message as in Fig.~\ref{message} (left), before sending it to the IOTA Tangle.     
  
  \item \textbf{Resolve} function is responsible for resolving a DID, that is, extract the corresponding DID Document from the IOTA Tangle while verifying the validity of the OTT message carrying the DID Document.

  \begin{figure}[h!]
    \begin{center}
    \includegraphics[width=7cm]{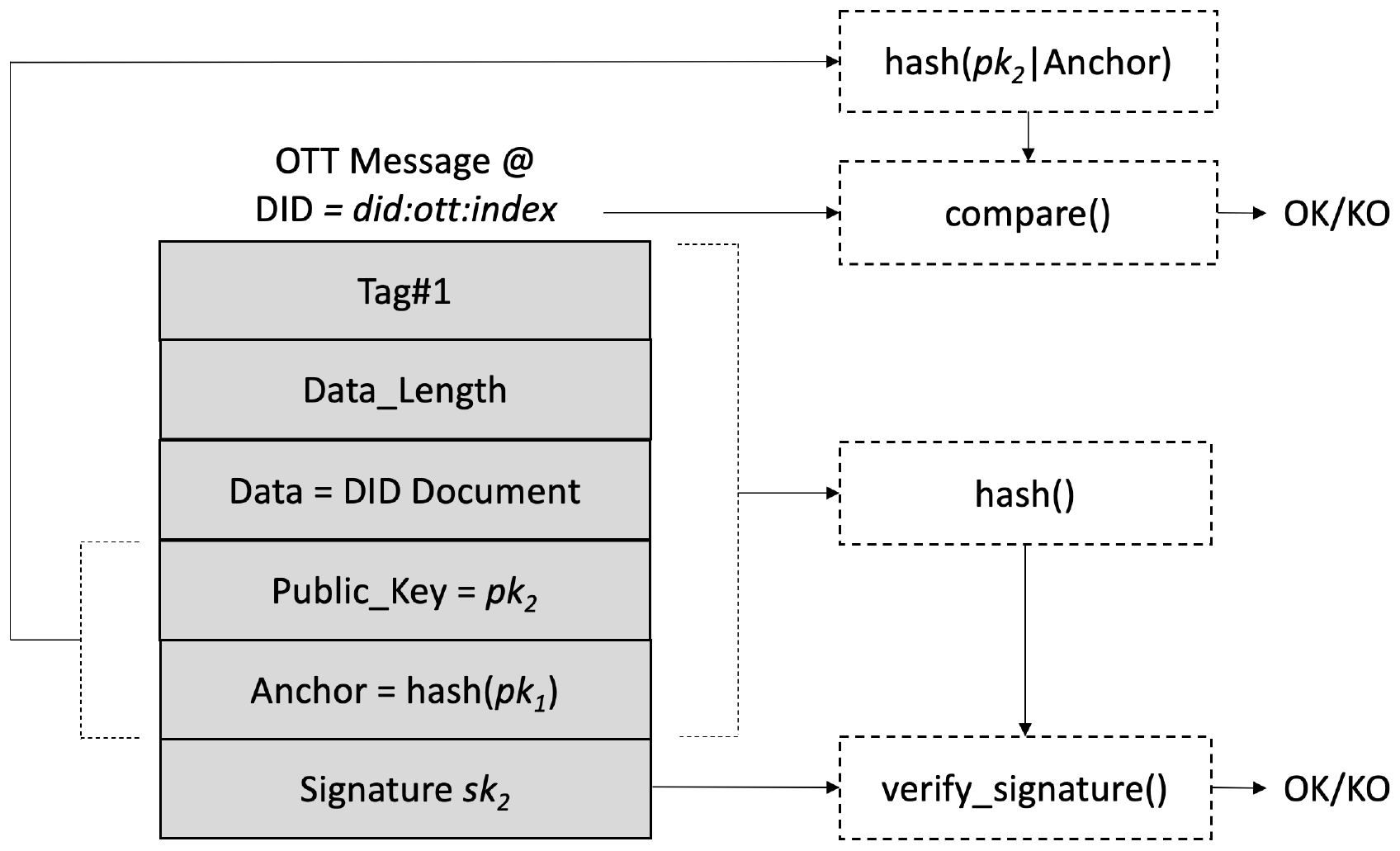}     
    \end{center}
        \vspace{-0.5cm}
    \caption{Validation of an OTT create message.} 
    \label{resolve_msg}
    \begin{center}
    \includegraphics[width=10cm]{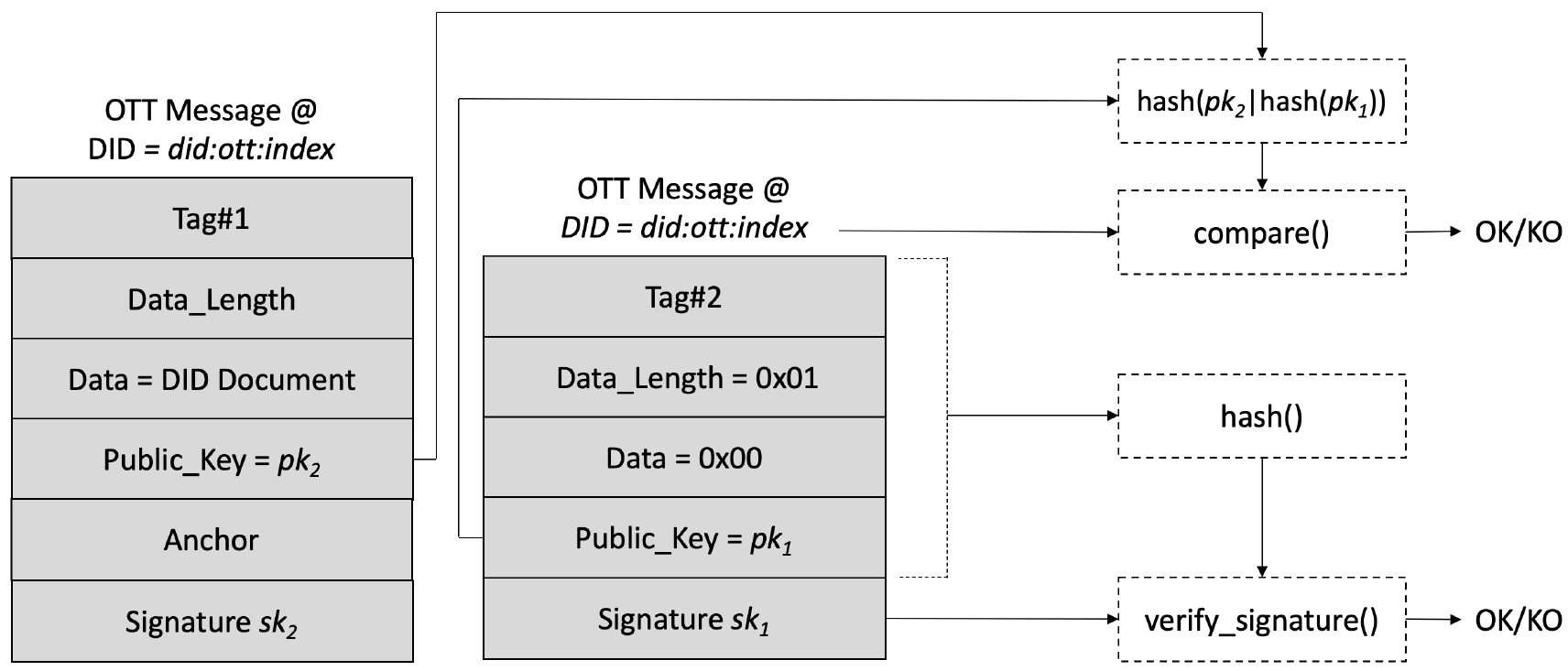} 
    \end{center}
        \vspace{-0.5cm}
    \caption{Validation of an OTT revoke message.} 
    \label{resolve_revoked}
        \vspace{-0.7cm}
\end{figure}

  In case of an OTT create message, the function checks its integrity by verifying the integrity signature field against the hash of the rest of the message as depicted in Fig.~\ref{resolve_msg} and, then, it checks that the DID has been generated in accordance with the OTT method, \ie\ $hash(pk_2|Anchor) \equiv index.$
  
  In case of an OTT revoke message, the function checks its integrity by verifying the signature field against the hash of the rest of the message as depicted in Fig.~\ref{resolve_revoked} and, then, it checks whether the peer who revoked the DID is the same who created it by verifying that the relations among the public key $pk_1$ in the revoke message, $pk_2$ in the create message, and the DID holds: $Anchor \equiv hash(pk_1) \land hash(pk_2|hash(pk_1)) \equiv index$.

  The Resolve function returns the DID Document in case of a valid DID, an empty document in case of a revoked DID and a proper error code in other possible cases.
  \item \textbf{Revoke} function is responsible for revoking a DID by means of an OTT revoke message as in Fig.~\ref{message} (right).
  \item \textbf{Update} function is responsible for updating the DID as a combination of a Revoke and a Create. The function revokes the current DID and then generates the new one from new random seeds (\eg\ $seed^{'}_1$ and $seed^{'}_2$). 
\end{itemize}

\section{Experimental setup and results}
\label{results}
The tests has been designed to assess the performances of OTT functions over a long period of time. A single target node is requested to interact with the Tangle on the IOTA mainnet to create, update and revoke its own DID Document and to resolve the DID of a peering node. The setup is pretty simple, the target node interacts with the Tangle through a gateway node~\cite{HORNET} integral part of the distributed ledger. The same tests have been performed by exploiting an available public node~\cite{IOTA-node} and a private node deployed and operated in our lab. Both nodes are synchronized with the IOTA mainnet and they maintain the same version of the Tangle by design. However, the private gateway node is configured to serve only the requests from the target node to evaluate the influence of competing requests on the overall performances of OTT. The overall test campaign lasted one month during which the four OTT functions have been tested 1000 times each. The target node runs on an Ubuntu 20 LTS machine with kernel 5.15, with Intel(R) Core(TM) i7-10510U CPU @ 1.80GHz 2.30 GHz, 24.0 GB RAM. %

\begin{figure}
\subfigure{\includegraphics[angle=360, width=6.3cm]{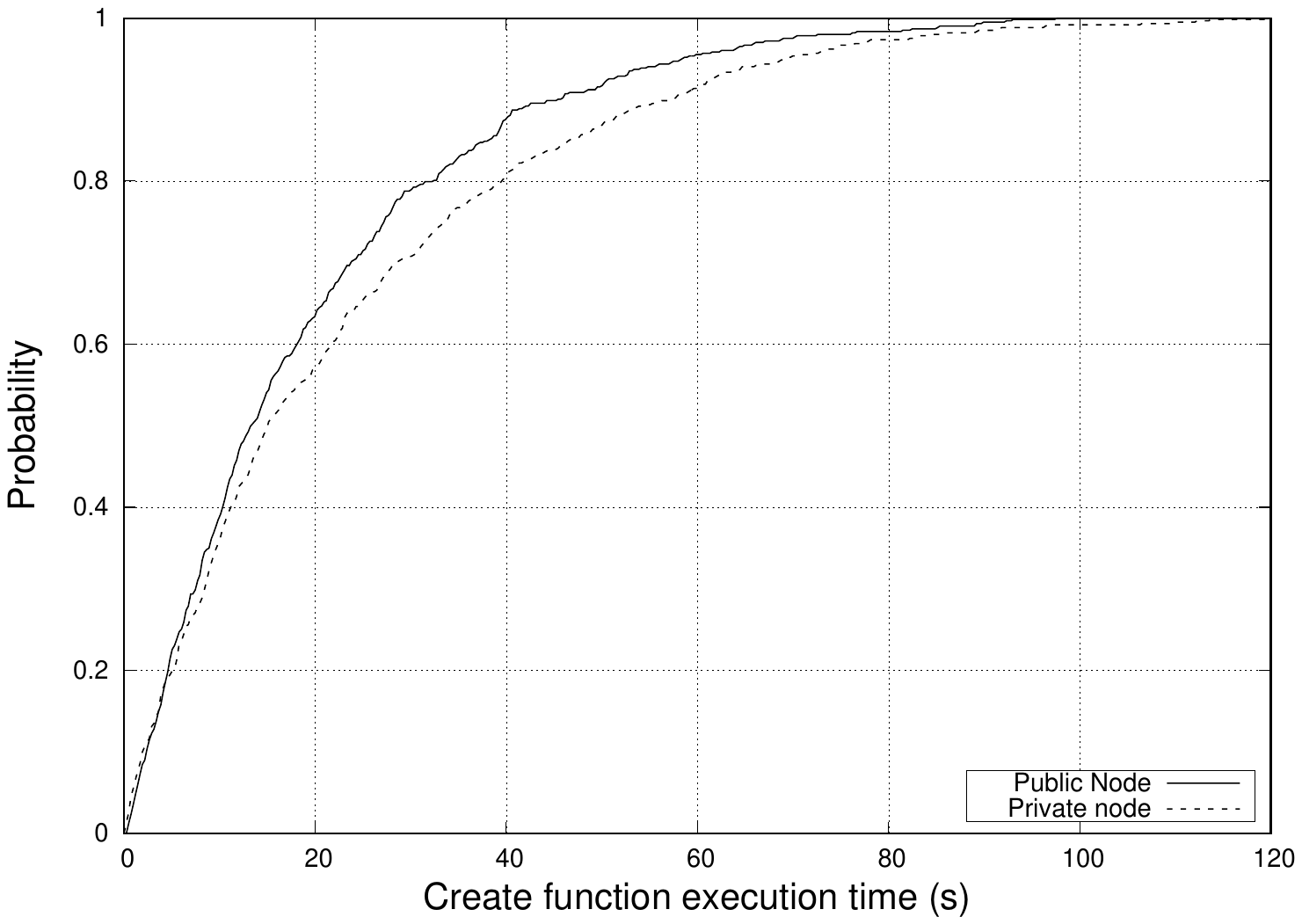}}
\subfigure{\includegraphics[angle=360, width=6.3cm]{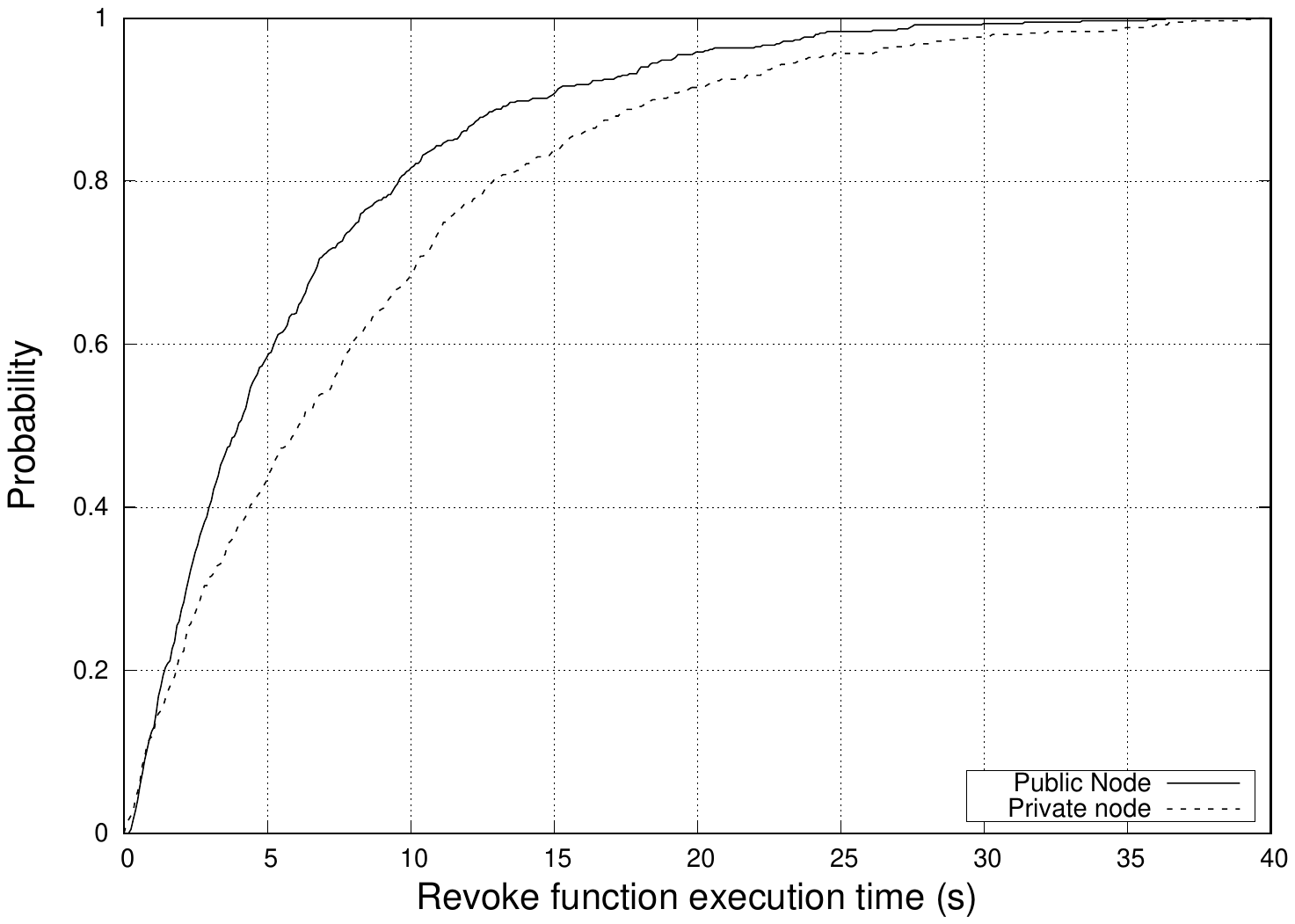}}
\subfigure{\includegraphics[angle=360, width=6.3cm]{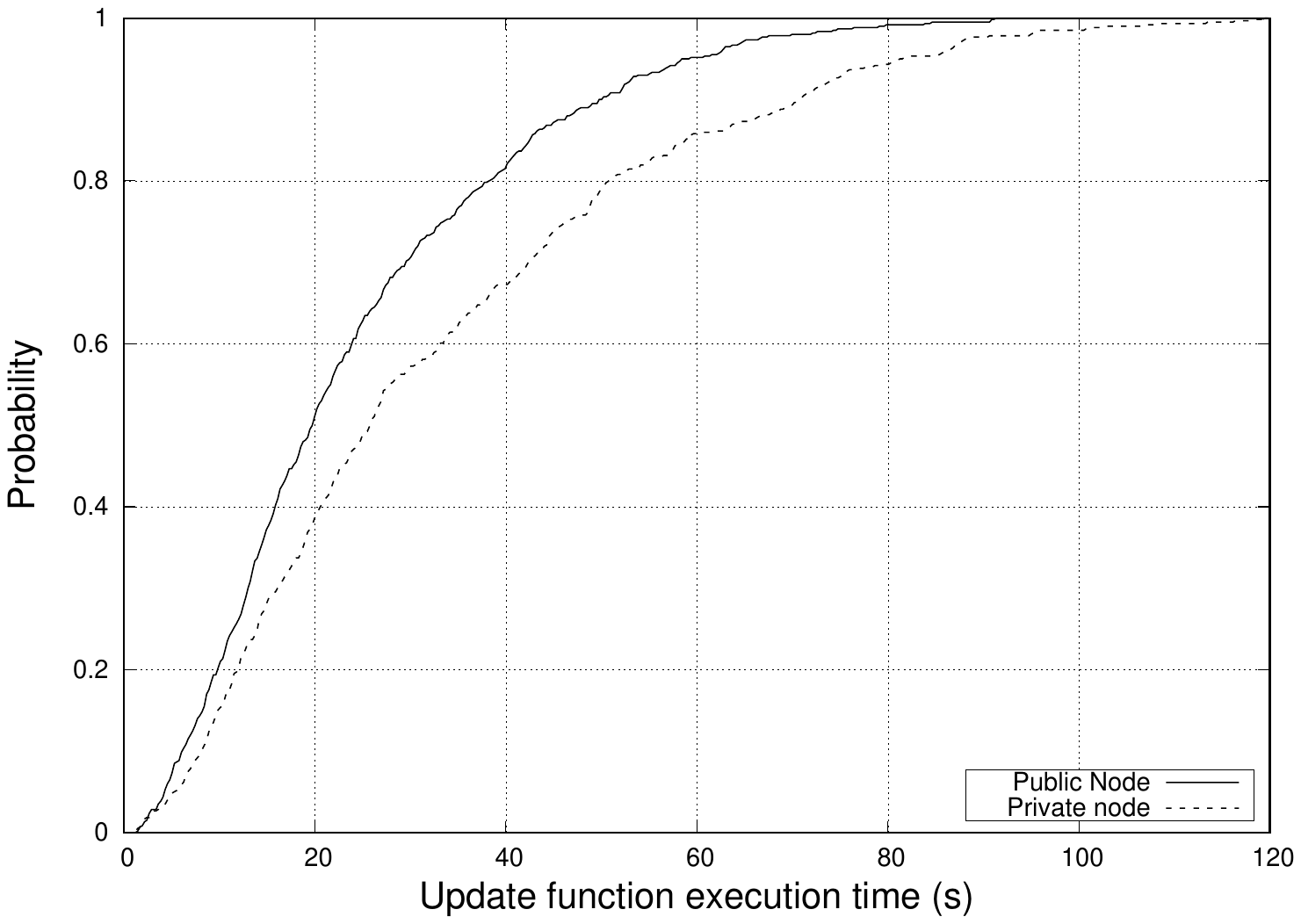}}
\subfigure{\includegraphics[angle=360, width=6.3cm]{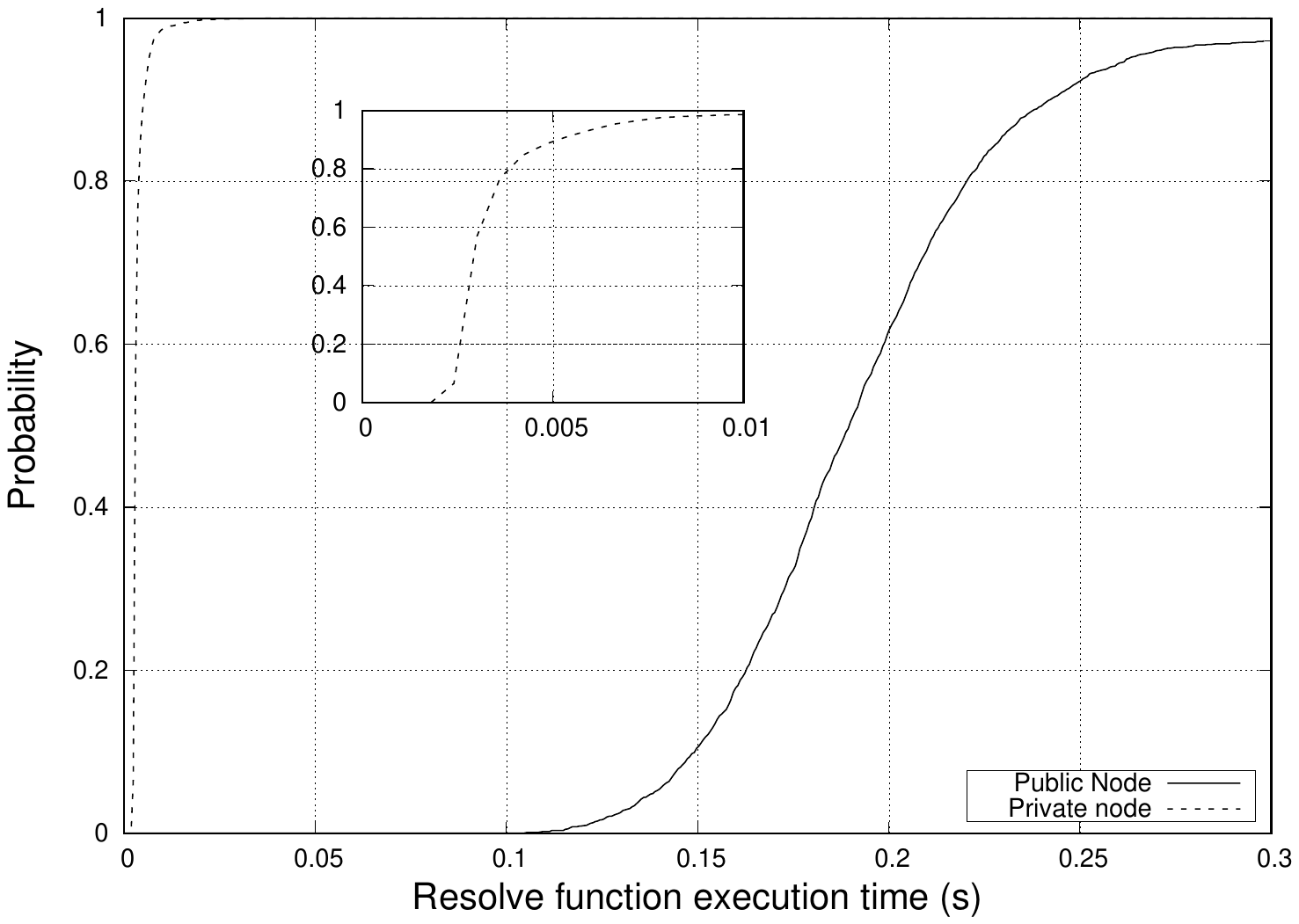}}
\caption{Empirical CDF of the execution time of the four OTT functions; figure in figure in the Resolve graph shows a zoom of the empirical CDF with the private node setup.}
\label{result-figures}
\vspace{-0.4cm}
\end{figure}

The graphs in Fig.~\ref{result-figures} show the empirical cumulative distribution function (CDF) of the execution time of the four OTT functions. %
Each value $y$ represents the CDF at $x$ of the empirical distribution calculated from the tests, \ie\ $F_T(t) = P(T \le\ t)$ where $t$ is the overall execution time of the specific OTT function. Each graph depicts the empirical CDF obtained with the public and the private gateway node for comparison. The empirical CDFs of the \emph{Create} function are comparable. 
The 0.95 quantile, \ie\ the estimated value such that 95\% of the execution times is less than (or equal to) that value, is equal to 58.44 s with the public node and 68.02 s with the private node. Both are of the order of tens of seconds, as expected. 
The \emph{Create} function lasts 19.47 s and 23.16 s in average with the public and the private node respectively. The same considerations stand for the \emph{Revoke} and \emph{Update} functions. 
A \emph{Revoke} last 6.03 s and 8.19 s with the public and private gateway node, respectively, whereas an \emph{Update} needs 24.25 s and 32.82 s by using the public and private gateway node respectively. The sum of the average time for the execution of a \emph{Create} and a \emph{Revoke} is about the same of an \emph{Update}, as expected by design of these functions.  
These results suggest that competing requests at a public gateway node do not influence the performances of the OTT functions that require a write on the IOTA Tangle. It is worth noting that the creation, the revoke and the update of a DID are performed only once or periodically at a low frequency, hence the choice on the adoption of a public or a private gateway node has a limited impact on the overall performance.
The empirical CDFs of the \emph{Resolve} function, shows, as expected, that the performance with the private gateway node outperforms the one with the public gateway node. Resolving a DID by leveraging an unloaded private gateway node is about one hundred times faster than with a public node. Resolving the same DID lasts 3.49 ms and 216 ms by querying the private and the public gateway node, respectively. This important result opens up new perspectives.

\section{Future Perspectives}

Most of the discussions about the DID-based authentication in the SSI world consider an implementation at the application layer of the TCP/IP stack. The primary reason for this choice is that most of the community comes from the WWW. The Authors of this paper believe in the implementation of the DID-based authentication at the transport layer of the TCP/IP stack through and directly within the Transport Layer Security (TLS) v1.3~\cite{rfc8446}. 
Implementing this option requires to modify the current TLS handshake to work with DIDs and DID Documents. %
This paper proposes to integrate the DID technology into OpenSSL, one of the most adopted cryptographic library, and made it transparently available in LibSSL for the adoption by the TLS protocol. The very fast execution of a DID resolution with the private setup, shown in Fig.~\ref{result-figures}, is an important result and it supports the proposed option. In fact, a DID-based TLS handshake skips the time for public key certificate chain verification (\ie\ the most computational demanding operation) but it pays a DID resolution, hence a fast DID resolution is desirable.
As a first step toward the adoption of DID technology into the TLS handshake, it is necessary to provide OpenSSL the implementation of the DID Method. The paper here presents for the first time, the design and implementation of a DID Provider specifically conceived to ease the integration of DID methods into OpenSSL, \ie\ makes it pluggable and avoids code refactoring. 

The integration of DID technology into OpenSSL begins with the definition of a novel Operation called DID. As shown in the code listing below, the definition of the DID Operation requires three main steps. First, assigning an integer value to the Operation through the identifier prefixed by the \emph{OSSL\_OP\_} string. Second, declaring the required template functions of the Operation through \emph{OSSL\_FUNC\_*\_*} strings that univocally defines the function's identifiers. Finally, using the macro \emph{OSSL\_CORE\_MAKE\_FUNC}() to define the OpenSSL name of the functions (\ie\ \emph{did\_create, did\_update, did\_resolve} and \emph{did\_revoke}), the input and the return values. Developers calls those functions from within OpenSSL through a set of APIs.

\begin{verbatim}
#define OSSL_OP_DID             24
#define OSSL_FUNC_DID_CREATE    1
#define OSSL_FUNC_DID_RESOLVE   2
#define OSSL_FUNC_DID_UPDATE    3
#define OSSL_FUNC_DID_REVOKE    4
OSSL_CORE_MAKE_FUNC(void *, did_create, 
    (void *sig, size_t siglen, int type))
OSSL_CORE_MAKE_FUNC(int, did_resolve, 
    (char * index, DID_DOCUMENT* did_doc))
OSSL_CORE_MAKE_FUNC(int, did_update, 
    (char * index, void *sig, size_t siglen, int type))
OSSL_CORE_MAKE_FUNC(int, did_revoke, (char * index))
\end{verbatim}    

This DID Operation definition, allows the developers of DID Methods to make them available through OpenSSL via a Provider in an \emph{agile} fashion. The term agile means that developers are provided with a stable logic to simply plug their own specific implementations and make them available in OpenSSL without major code refactoring. 
Thus, an OpenSSL application can load the Provider offering the DID Operation and ask it to supply the implementation of the chosen DID Method functions. 
In case of a DID Provider offering multiple different DID Method implementations, \eg\ DOM for Ethereum, and OTT for IOTA Tangle, as shown in the listing below:
\begin{verbatim}
operation_did[] = {
 {"DOM","provider=didprovider", didprovider_dom_functions},
 {"OTT","provider=didprovider", didprovider_ott_functions}}
\end{verbatim}
an application can request OTT method by calling \verb+DID_fetch(didctx,"OTT")+
where \verb+didctx+ is the OpenSSL context, to receive the pointers to the implementations of the \emph{did\_create, did\_update, did\_resolve} and \emph{did\_revoke} functions of the OTT method at run time. 

Finally, OTT method has been plugged into OpenSSL via the DID Provider and the new tests to evaluate the overhead introduced by OpenSSL when calling the OTT functions showed no significant statistical variation with respect to the ones presented in Section~\ref{results}; hence they are not reported here for conciseness.

\section{Conclusions and Future Works}
The paper has presented, for the first time, a novel DID Method called OTT and has discussed its design and working principles that leverage the IOTA Tangle as the RoT for identity data. 
The results of a long lasting experimental test campaign in real-world settings has suggested the adoption of a private gateway node synchronized with the IOTA Tangle on the mainnet for efficient DID control, in particular for reducing of about 100x the time for a DID resolution. This result has suggested the Authors to deal with the integration of DID technologies into OpenSSL and to start addressing DID-based TLS handshake in SSI framework while maintaining the interoperability with public key certificates. The results of the implementation of a DID-based TLS v1.3 handshake and its security analysis will be provided in a future paper.

\subsubsection{Acknowledgements} 

This work has been developed within the SEDIMARK project \emph{https://sedimark.eu/}. SEDIMARK is funded by the European Union under the Horizon Europe framework programme [grant no. 101070074]
\\

This preprint has not undergone peer review or any post-submission improvements or corrections. The Version of Record of this contribution is published in Blockchain and Applications, 5th International Congress, and is available online at https://doi.org/10.1007/978-3-031-45155-3\_38

\bibliographystyle{splncs04}
\bibliography{biblio}
\end{document}